# Probe Skyrmion phases and dynamics in MnSi via the magnetoelectric effect in a composite configuration

Yisheng Chai[1,2,*], Peipei Lu[3,4†], Haifeng Du[5], Jianxin Shen[4], Yinina Ma[4], Kun Zhai[4], Le Wang[4], Youguo Shi[4], Hang Li[4], Wenhong Wang[4], and Young Sun[1,2,4‡]

[1] Center of Quantum Materials and Devices, Chongqing University, Chongqing 401331, China

[2] Low Temperature Physics Laboratory and Chongqing Key Laboratory of Soft Condensed Matter Physics and Smart Materials, College of Physics, Chongqing University, Chongqing 401331, China

[3] College of Physics, Hebei Normal University, Shijiazhuang 050024, Hebei, China

[4] Beijing National Laboratory for Condensed Matter Physics, Institute of Physics, Chinese Academy of Sciences, Beijing 100190, China

[5] High Magnetic Field Laboratory, Chinese Academy of Science, Hefei 230031, Anhui, China

[*] yschai@cqu.edu.cn

[†] peiplu@hebtu.edu.cn

[‡] youngsun@cqu.edu.cn

## Abstract

We have developed a sensitive technique to probe the magnetic skyrmion phases and dynamics by employing the interfacial coupling effect in a magnetoelectric composite configuration. The study on a MnSi single crystal sample using this technique provides clear evidences for the skyrmion lattice phase and coexistence of skyrmion and conical phase. Above the Curie temperature $T_C$, a region with strong spin fluctuation is revealed as well. By tuning the density of Skyrmion or disorder, a transition from the skyrmion lattice to skyrmion-conical coexisting phase is observed. The observation is in good agreement with a theoretical model which predicts the dissipation behavior in the coexistence phase.

## I. INTRODUCTION

A wide class of condensed matter systems can be effectively described as a collection of interacting quasi-particles, like the vortices in type-II superconductors, charge density waves, Wigner crystals of electrons and Skyrmion (**Sk**) phase in chiral magnets, by forming a hexagonal lattice of particles[1-4]. Skyrmion, a topologically protected quasiparticle in magnets [5,6], is attracting a great deal of attention due to its intriguing physics [7-10] as well as the possible applications in next generation memory devices [11-16]. On one hand, the closely packed hexagonal two dimensional (2D) lattice of Sks has been well visualized in the non-centrosymmetric chiral magnets such as MnSi [4,17,18] and $Cu_2OSeO_3$ [19-22] with hundreds of nm size by Lorentz Transmission Electron Microscopy. On the other hand, the existence of **Sk** phase in bulk sample can be well identified by many existing tools, like magnetoresistance, Hall, magnetization, ac susceptibility, Neutron diffraction, ultrasonic sound *etc* [18,23-27]. However, a method to clearly distinguish between pure lattice and coexistence phases in bulk sample is still lacking. Need to point out that, the condition to stabilize the **Sk** phases can be very different between bulk and nano-scaled samples [25,28]. In this sense, the microscopic image is not suitable for this purpose.

Here, we show that the technique of piezoelectric transducer allows the probing of ac magnetoelastic response of **Sk** phases in bulk samples via interfacial strain coupling in a composite magnetoelectric (ME) configuration. In an ME composite, it consists of both magnetostrictive and piezoelectric phases in which the mechanical deformation due to magnetostriction results in a change of polarization *P* due to piezoelectric effects [29]. Based on previous studies, the **Sk** phase in MnSi is found to show clear field dependent magnetostrictive behavior [30]. Therefore, a Sk bulk sample (magnetostrictive phase) can be mechanically bonded with a thin piezoelectric PMN-PT ($0.7Pb(Mg_{1/3}Nb_{2/3})O_3$–$0.3PbTiO_3$) transducer (200 μm thick, piezoelectric phase) with the interfacial strain-mediated composite configuration [29], as schematically shown in Fig. 1(a). Then, its ME response is the presence of *P* upon applying a magnetic field (*H*):

$$\Delta P = \alpha \Delta H \text{ or } \Delta E = \alpha_E \Delta H \quad (1)$$

where *E* denotes electric field and α ($\alpha_E$) is the ME (ME voltage) coefficient. According to the previous theoretical studies, the longitudinal ME voltage coefficient $\alpha_{E33}$ ($=dE_3/dH_3$) of ME laminates can be roughly described by the following relationship

[29]:

$$\alpha_{E33} \propto kfd_{31}q_{31} \tag{2}$$

where $f$ is the volume fraction of magnetic phase, $d_{31}$ the piezoelectric coefficient, the interface strain coupling parameter $0 < k < 1$, $q_{31}$ (=$d\lambda_1/dH_3$) is the magnetostrictive coefficient where $\lambda_1$ is the magnetostriction along **1** direction under external $H$ along **3** direction. This means that the ac magnetostrictive response $q_{31}$ of the **Sk** phase can be converted into $\alpha_{E33}$ and studied electrically with a very high sensitivity by lock-in technique.

In this letter, we found that $\alpha_{E33}$ of the MnSi/piezoelectrics laminate has a much higher sensitivity in all the magnetic phases in MnSi than the traditional techniques such as ac magnetic susceptibility, and thus enables us to map out the phase diagram with distinct dynamic characteristics. $\alpha_{E33}$ shows strong out-of-phase components in the **Sk**-conical (**Sk-C**) coexistence region in the single crystal MnSi sample, indicating the dissipative motion of **Sk-C** domain boundary under ac driven magnetic fields. Surprisingly, the out-of-phase component of $\alpha_{E33}$ only disappears deep inside the **Sk** phase, revealing the stabilization of pure Sk lattice via the strong Sk-Sk interaction and low density of pinning centers. Above $T_c$ = 29.6 K, our technique reveals a short-range correlation region below paramagnetic (**PM**) phase as well.

## II. MATERIALS AND METHODS

Single crystals of MnSi were synthesized using a Ga self-flux method in alumina crucible, which was sealed in a fully evacuated quartz tube. The crucible was heated to 1150 °C in 12 h and dwell for 20 h, then cooled slowly to 950 °C at 2 K/h, where the Ga flux was spun off in a centrifuge. The residual Ga flux on the surface of crystals was resolved in concentrated nitric acid. High quality single crystals were obtained by washing with water repeatedly. The typical sample geometry is 2 × 1 × 0.5 $mm^3$ with the widest face perpendicular to [110] direction. Magnetic ac susceptibility $\chi=\chi'+i\chi''$ were measured by a Magnetic Properties Measurement System (MPMS-XL, Quantum Design) under various frequencies with ac magnetic field $H_{ac}$ = 1 Oe. To prepare the MnSi/PMN-PT ME laminate, first, one side of MnSi was polished with less fine sand paper, and a commercially bought [001]-cut PMN-PT single crystal with 0.2 mm thickness was attached on the surface with silver epoxy (Epo-Tek H20E, Epoxy Technology Inc.). The other side of the PMN-PT is also covered with silver epoxy to measure the electrical signal of PMN-PT due to piezoelectric effect. The MnSi/PMN-PT bonded with silver epoxy constitutes an ME laminate device, as shown in Fig. 1(a). Since the thickness of the MnSi sample is much larger than that of the PMN-PT single crystal, the induced distortion by sample is regarded to be homogeneous for PMN-PT. The layers of the ME composite are oriented along the planes (**1**,**2**) and that the **3** axis is perpendicular to the same plane. In this case, the direction of polarization in PMN-PT coincides with the **3** axis, as shown in Fig. 1. Before the ME voltage coefficient measurements, the MnSi/PMN-PT laminate was poled by applying a dc electric field of 650 kV/m from a Keithley 6517B electrometer at room temperature. Then, the longitudinal ME voltage coefficient $\alpha_{E33}=dE_3/dH_3$ were measured in a Cryogen-free Superconducting Magnet System (Oxford Instruments, Teslatron PT) using a home-made sample probe, as shown in Fig. 1(b). The dc magnetic field $H_{dc}$ is generated and swept by a superconducting magnet. A Keithley 6221 ac source was used to supply an ac current to the helmholtz coil to generate ac magnetic field $H_{ac}$ = 1 Oe in a broad frequency range (33 Hz – 1 kHz) and the resultant ac voltage $V_3 = V_x + iV_y$, across the electrodes was measured by a lock-in amplifier (Stanford Research SR830) as a function of $H_{dc}$ or temperature. The ac ME voltage coefficient $\alpha_{E33}$ is calculated by $\alpha_{E33} = (V_x + iV_y)/(H_{ac}t)$, where $t$ is the thickness of the ME composite. (Fig. 1(b)). Represented as $\alpha_{E33} = \alpha_x + i\alpha_y$, the $\alpha_{E33}$ is composed of in-phase component $\alpha_x$ and out-of-phase component $\alpha_y$.

## III. RESULTS AND DISCUSSION

Fig. 2(a) shows the $H_{dc}$ dependence of the ac susceptibility χ' and χ'' for $H_{dc}$//[110] direction at 28 K with various frequencies. There are clear features which correspond well with the expected magnetic phases and phase transitions [16,18, 26], in particular, proving the existence of the **Sk** phase at this temperature. For high magnetic field $H$ > 3.8 kOe, MnSi is in the ferromagnetic (**F**) phase. With $H$ down swept from 3.8 to 2.4 kOe, χ' shows a drastic increase, indicating an **F** to the conical (**C**) phase transition. Further decreasing of $H$ from 2.4 to 0.4 kOe, χ' displays a clear dip in the middle of the field region, which indicates the **Sk** phase sandwiched between two **C** phases. Between 0.4 to 0 kOe, there is another dip in χ', marking the helical (**H**) phase centered at zero $H$. All the above features in χ' are fully consistent with that in literature [26]. In the case of χ'', the two phase-boundaries between **C** and **Sk** are marked by two peaks while there is very weak $H$ dependent behavior for other phases or phase transitions. There are very clear frequency dependent behaviors around the **Sk** to **C** phase boundaries. From the above discussions, we could determine the successive transition fields $H_{c1}$, $H_{a1}$, $H_{a2}$, $H_{c2}$ of **H** to **C**, **C** to **Sk**, **Sk** to **C** and **C** to **F** phase transition respectively in the $H$ decreasing run, as exemplified in Fig. 2(a) for $T = 28$ K.

On the other hand, the MnSi sample is mechanically bonded with a PMN-PT [001]-cut thin plate to form an ME composite. We measured the $\alpha_{E33}$ of the MnSi/PMN-PT laminate at selected temperatures with $H_{dc}//E$ configuration, as shown in Figs. 2(b)-2(d). The strong field dependent $\alpha_{E33}$ indicates a strong magnetoelastic behavior in the magnetic phases of MnSi and a good interfacial bonding condition between MnSi and PMN-PT. At 28 K, each magnetic phase indeed generates distinctive ME signals that correspond well to the field values in magnetic susceptibility at this temperature, as shown in Fig. 2(b). In particular, the **Sk** phase is marked by clear features in both $\alpha_x$ and $\alpha_y$, confirming the validity of this method.

Figs. 2(c) and 2(d) further show the field dependent $\alpha_x$ and $\alpha_y$ at selected temperatures of the sample. All the curves are nicely centrosymmetric to the original point, consistent with that of the conventional ME composites [29]. The magnitude of $\alpha_x$ is always much larger than that of $\alpha_y$ at high $H$, indicating little or negligible phase lag from the measurement system. At 20 K, $\alpha_x$ and $\alpha_y$ only show a small wiggle around ±0.5 kOe, corresponding to the **H** phase. Between 1 kOe and 5 kOe, $\alpha_x$ and $\alpha_y$ are almost constant and very close to zero value, which is attributed to the **C** phase. A sharp enhancement is observed around 5.3 kOe for $\alpha_x$ while a small wiggle is observed in $\alpha_y$, indicating the metamagnetic transition from **C** to **F** phase. However, for temperatures above 26 K, both $\alpha_x$ and $\alpha_y$ gradually show peak-dip features in the field range of 1 to 2.5 kOe. These peak-dip features become more pronounced with increasing temperature, and persist up to 28.6 K. Above 28.6 K, the entire $H$ dependent $\alpha_x$ and $\alpha_y$ curves are very smooth and only show a broad peak around 3.5 kOe. With further increasing the temperature, the ME signal slowly decays above $T_c$, indicating a strong short-range correlation between spins in this temperature region. By referring to the reported phase diagram of bulk MnSi [16,18,26], the appearance of the peak-dip features in the intermediate field regime clearly points to the expected magnetic **Sk** phase. We also see very clear frequency dependent $\alpha_x$ and $\alpha_y$ at **Sk** to **C** phase boundaries, consistent with the magnetic susceptibility data at the same frequencies.

To further reveal the details of the observed **Sk** phase in MnSi, we converted the field dependent complex $\alpha_{E33}$ data (Figs. 2(c) and 2(d)) into the Argand diagram (using $\alpha_x$ as the real axis and $\alpha_y$ as the imaginary axis in the complex plane), as shown in Fig. 3(a). By using the Argand diagram, a dissipative magnetic phase that contributes to an out-of-phase component in $\alpha_{E33}$ is clearly distinguished. The Argand plot of $\alpha_{E33}$ at 30 K (>$T_c$), which is in the **PM** phase region, shows a straight line with a fixed small phase angle in the entire scanned field range of ±6 kOe. This means that there is no dissipative behavior in **PM** phase under the ac driven field. The small phase angle should come from the measurement system. Below $T_c$, all other phases except the **Sk** phase lie on the same baseline with the same phase angle, marking the non-dissipative nature of the **H**, **C** and **F** phases in MnSi. In contrast, the Argand diagram within the **Sk** phase shows a three-section feature (I-II-III), as exemplified at 28 K in Fig. 3(b). The curves in I (1.0 to 1.5 kOe) and III (2.0 to 2.4 kOe) sections show negative and positive out-of-phase components with respect to the base line, respectively, under continuous variation of $H$, indicating the dissipative behaviors within these two sections. However, the intermediate II (1.5 to 2.0 kOe) section resides on the same baseline with other magnetic phases, pointing to a non-dissipative nature of this section. At lower $T$, the out-of-phase components of the I and III sections gradually shrink together and collapse towards the baseline. The II section disappears at 27 K below which only the I and III sections exist. At higher $T$ region between 28.6 K and $T_c$, the I and III sections gradually collapse onto the base line with increasing temperature, maybe due to the strong thermal fluctuation near $T_c$.

From the above data, the I and III sections, *e.g.* at 28 K, should be attributed to the expected **Sk-C** coexistence phase coexistence region, as indicated in Fig. 3(c). As the I and III sections are directly connecting with the **C** phase, they have relatively lower density of **Sk** than that of the II section. Dissipations would be expected during the nucleation of the skyrmions, and the motion of the domain walls separating the **C** phase from the **Sk** phase. On the contrary, with increasing the Sk density from the I or III section to the II section which has the highest Sk density, the stronger Sk-Sk interaction enhances the elastic properties significantly, leading to a condensation of the Sk-lattice in section II. The low density of pinning centers in a single crystal should not break the regularity of Sk-lattice, as shown in Fig. 3(d). More importantly, the coherent motion of Sk-lattice is less disturbed under the ac driven field due to the finite elastic properties in the lattice, leading to a negligible phase lag in $q_{31}$, and consequently $\alpha_{E33}$. The non-dissipative behavior in the II section strongly suggests the realization of the pure Sk-lattice phase. The strong dissipating behavior of **Sk**-**C** coexistence region under large $H_{ac}$ can be qualitatively understood by a classical model of the forced small harmonic oscillation with damping [30,31]. Note that, in the Figs. 2(b) and 2(d), the $\alpha_y$ at 28 K in the mixed phase region shows negative values. This is more clearly demonstrated in the I section of Fig. 3(b) where **Sk** and **C** phases coexist. In contrast, imaginary part of the magnetic susceptibility, which is related to the dissipation, is positive all the time (Fig. 2(a)). This controversy in I section can be resolved in the phase coexisting scenario that **C** phase always contribute a larger positive $q_{31}$ without dissipation while the smaller $q_{31}$ of **Sk** phase in magnitude changes its sign from positive to negative during passing this section. The sign reversal of $q_{31}$ of **Sk** reverses its imaginary part as well. As a result, the according real part of $\alpha_{E33}$ does increase and decrease while the imaginary part of $\alpha_{E33}$ changes from positive to negative in this section.

When we summarize all the $H$ scan data at each $T$ and draw the phase diagrams of MnSi, a **Sk**-**C** coexistence phase appears to surround a Sk-lattice phase in the intermediate field region, as shown in Fig. 3(e). Therefore, by tuning the density of Sk, we indeed induce a Sk-lattice phase deep inside the **Sk**-**C** coexistence regions in the sample. To the **Sk** phase very close to $T_C$, its attribution of pure lattice or coexisting phase is unclear since the anomaly in $\alpha_{E33}$ of the **Sk** phase is too weak.

To further confirm the high resolution magnetic phase diagram in MnSi with the composite ME approach, we measured the temperature dependent $\alpha_{E33}$ of the MnSi/PMN-PT laminate under the selected $H_{dc}$, as shown in Figs. 4(a)-4(d). Under the dc magnetic field of 6.0 kOe, the sample is in the **F** phase below 32.5 K and there is only a broad peak for $\alpha_x$ at high temperature indicating the transition from **PM** to **F** phase. This broad peak above 30 K appears at every $\alpha_x$ curves. Below 4 kOe, this peak marks the boundary between **PM** and short-range correlation, which is dubbed as fluctuation disordered in literature [32]. There are another broad bumps for $\alpha_x$ under 4.0 and 5.0 kOe at 25.4 and 28 K respectively, consistent with the **C** to **F** phase transition. Under $H_{dc}$= 3.0 kOe, there is no extra feature in magnetoelectric signals where only C phase exists below 29.6 K. Under 2.5, 2.0 and 1.0 kOe, peaks or dips in $\alpha_x$ and $\alpha_y$ at 28.6, 28.8 and 29 K reveal the transition from **C** to **Sk-C** coexistence region. The upward/downward peaks for $\alpha_x$ and weak feature in $\alpha_y$ under 1.5 kOe indicate that the sample centers into pure Sk-lattice phase. Under 0.5 kOe, a clear downward peak for $\alpha_x$ is observed, indicating the transition of **H** to **C** phase at 29.2 K. The transition temperatures obtained by the temperature dependent data are consistent with that derived from magnetic field scan data, as shown in Fig. 3(e). The temperature dependent $\alpha_{E33}$ further validate the **Sk-C** coexistence region and pure lattice phase in the single crystal MnSi.

## V. CONCLUSIONS

To conclude, our results clearly demonstrate the Sk-lattice phase and all other magnetic phases, even spin fluctuation region in a MnSi single crystal sample by the magnetoelectric composite technique. We note that the simple and phase-angle-sensitive technique can directly probe the dynamic response of Sk organization in bulk materials. It can be widely applied to the study of the vortices in the type-II superconductors, spin density wave, and other exotic particle-organized phases with unusual dynamic properties.

## ACKNOWLEDGEMENTS

This work was supported by the National Natural Science Foundation of China (Grant Nos. 11674347, 11674384, 11974065, 51725104, 11774399, 11474330), Chongqing Research Program of Basic Research and Frontier Technology, China (Grant No. cstc2020jcyj-msxmX0263), Fundamental Research Funds for the Central Universities, China (2020CDJQY-A056, 2020CDJ-LHZZ-010, 2020CDJQY-Z006), Projects of President Foundation of Chongqing University, China (2019CDXZWL002), Science and Technology Project of Hebei Education Department (QN2021088), Hebei Normal University (Grant No. L2021B09). Y. S. Chai would like to thank the support from Beijing National Laboratory for Condensed Matter Physics. We would like to thank X. Y. Zhou and G. W. Wang at Analytical and Testing Center of Chongqing University for their assistance.

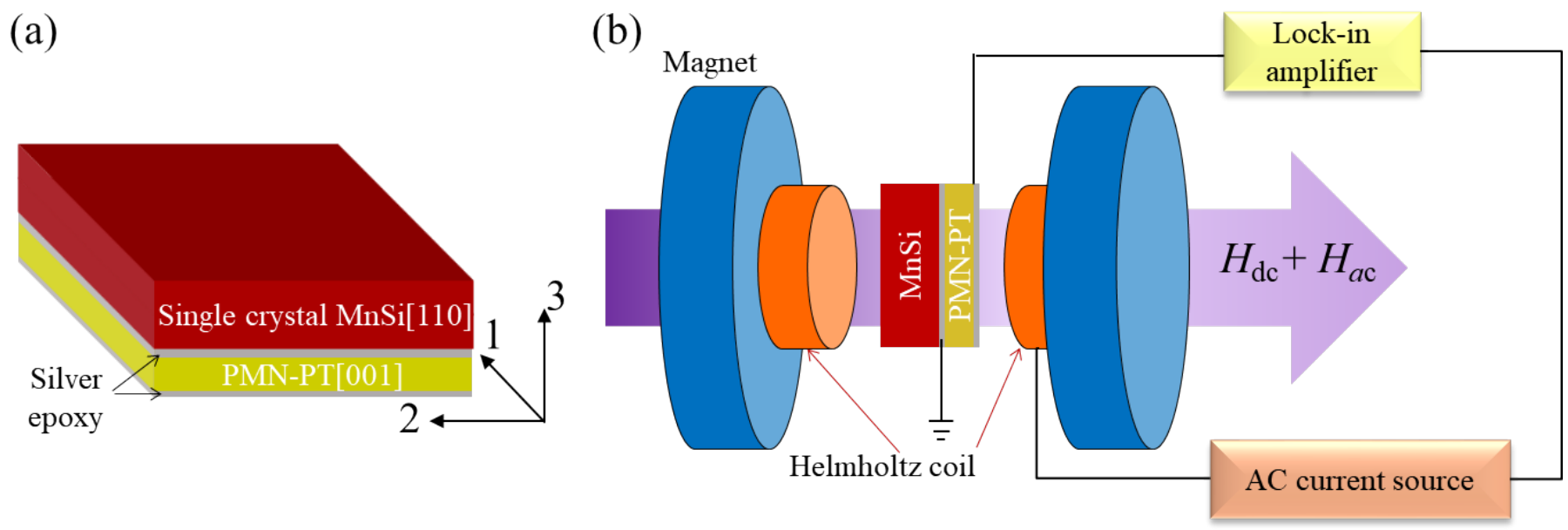

FIG. 1. (a) Schematic structure of the MnSi/PMN-PT laminate. (b) The schematic measurement system for the ac ME voltage coefficient $\alpha_{E33}$.

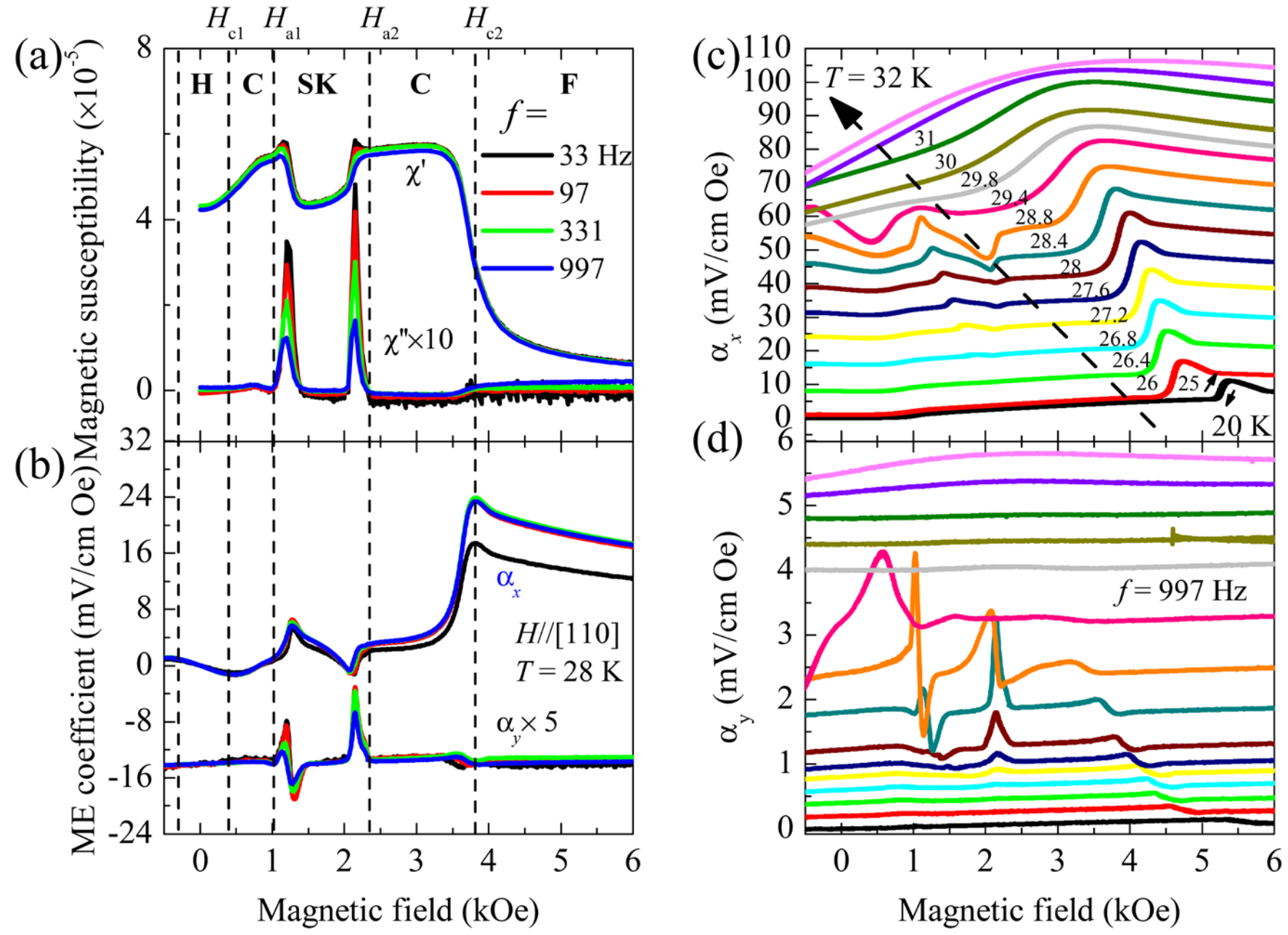

FIG. 2. (a) Magnetic field dependence of the ac magnetic susceptibility at various frequencies of 33, 97, 331 and 997 Hz at 28 K. (b) The ac ME voltage coefficient $\alpha_x$ and $\alpha_y$ at various frequencies as a function of magnetic field at 28 K. $\alpha_y$ is shifted vertically for clarity. (c) $\alpha_x$ and (d) $\alpha_y$ as a function of magnetic field at selected temperatures under the driven frequency of 997 Hz. They have been shifted vertically for clarity.

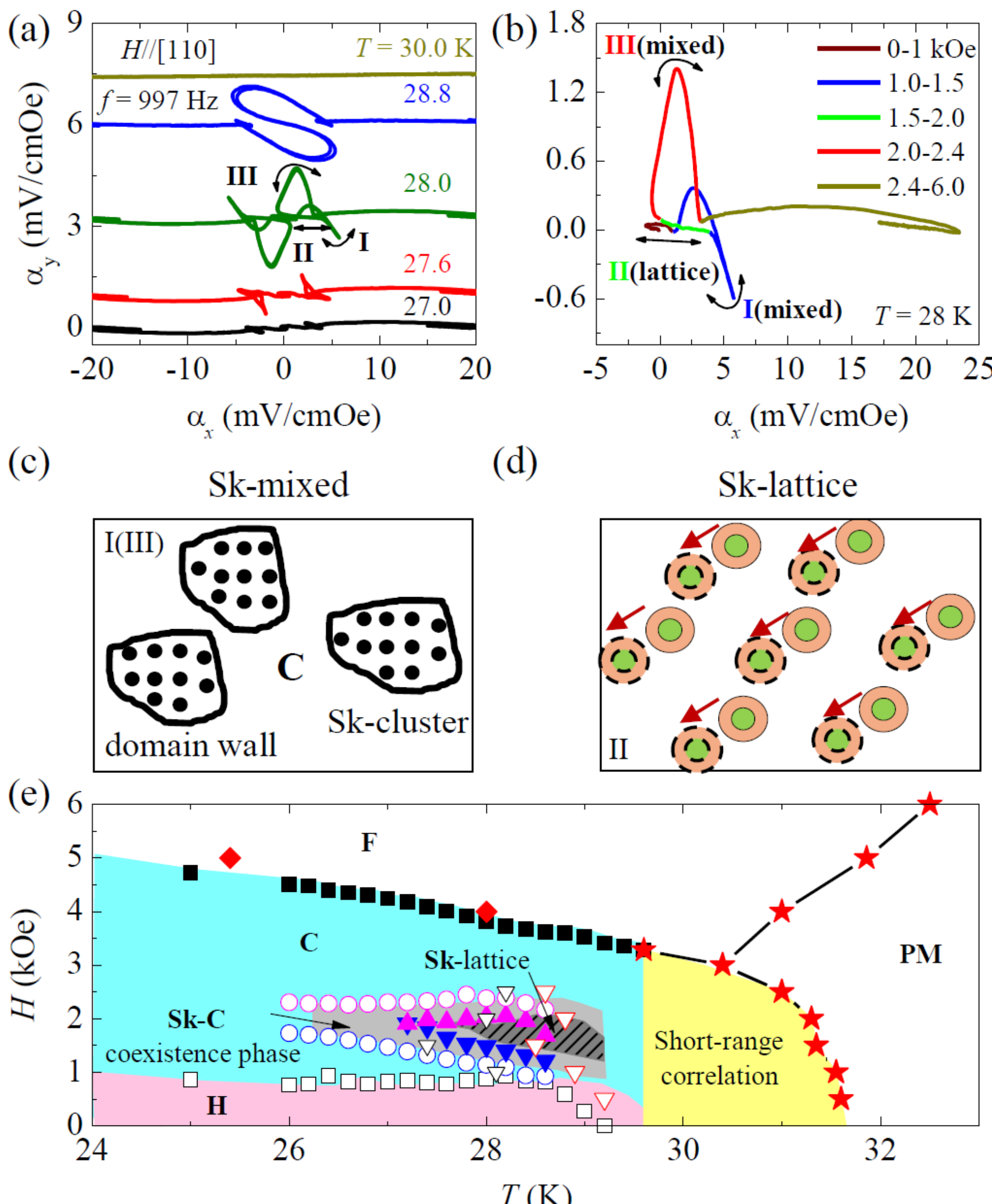

FIG. 3. (a) $\alpha_{E33}$ of a MnSi/PMN-PT laminate in the Argand diagram at selected temperatures with a frequency of 997 Hz. (b) The Sk lattice and mixed phase are marked by I, II and III sections at 28 K. Schematic illustration of the low-density (c) Sk-mixed and (d) Sk-lattice phases. (e) The phase diagram of the MnSi sample deduced from the Argand diagrams in (a) and Fig. 4.

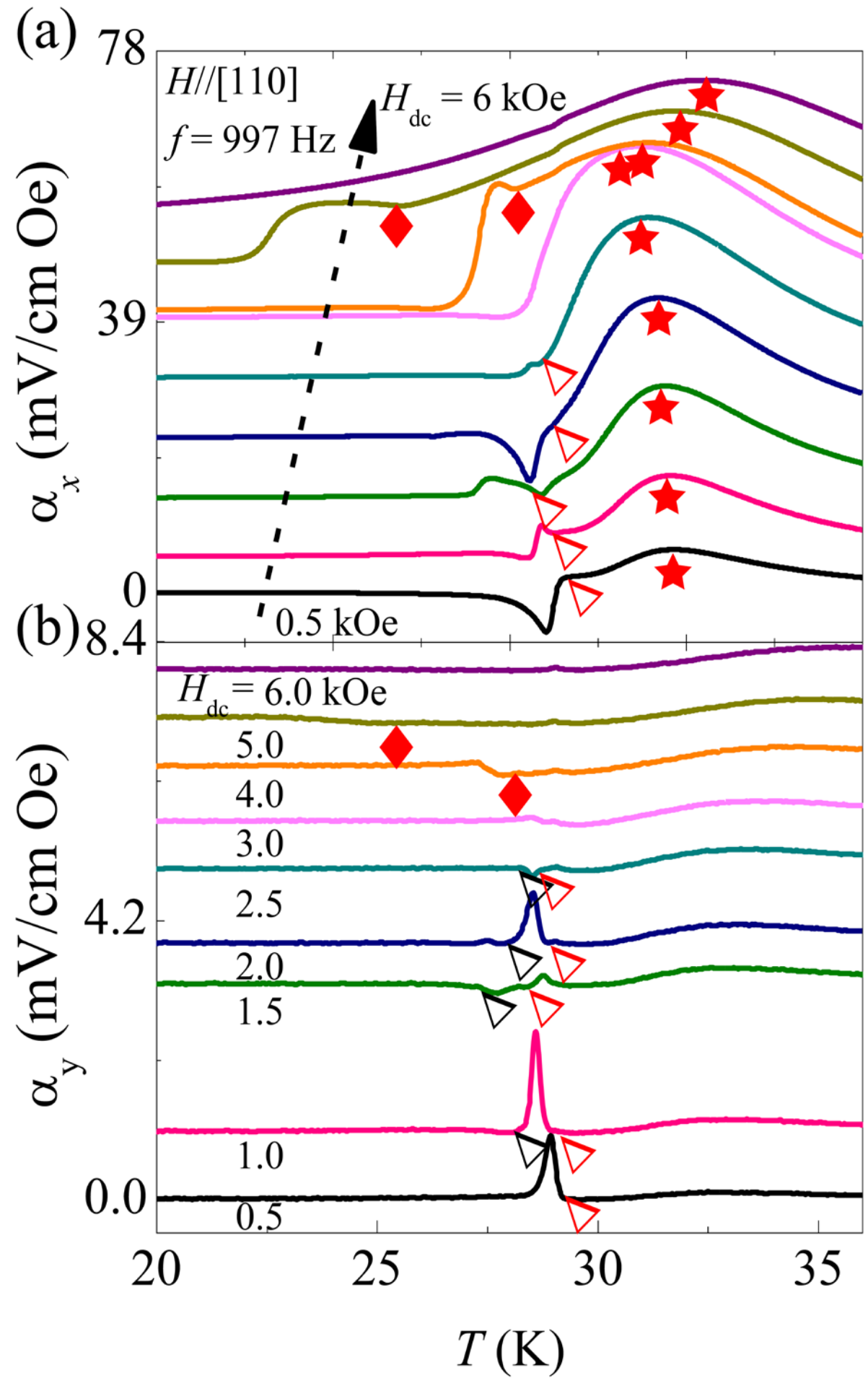

FIG. 4. Temperature dependence of (a) $\alpha_x$ and (b) $\alpha_y$ of MnSi/PMN-PT under selected $H_{dc}$.